# Memristor properties of high temperature superconductors

N.A. Tulina

Institute of Solid State Physics RAS, Chernogolovka

tulina@issp.ac.ru

**Abstract.**

*The review of studies on memristive properties or effect of resistive switchings in four classes of high temperature superconductors (HTSCs): $Bi_2Sr_2CaCu_2O_{8+y}$ (BSCCO), $YBa_2Cu_3O_{7-d}$ (YBCO), $Ba_{0.6}K_{0.4}BiO_{3-x}$ (BKBO), $Nd_{2-x}Ce_xCuO_{4-y}$ (NCCO) is presented in order to reveal functional properties of HTSCs which become apparent in the effects under discussion, prospects of usage of HTSC based memristors in applications and search for new mechanisms of strongly correlated nature to realize new generation memristors. The properties are: undergoing metal insulator transition at oxygen doping, transport anisotropy, existence of charge reservoirs through which doping of conductive copper oxygen plays is carried out. These are the main functional properties of HTSCs which permit to use them in memristors. By the example of study of bipolar effect of resistive switching in HTSC based heterojunctions it is shown how one can form memristor structures based on HTSCs using their functional properties.*

**Introduction**

Today it is considered that the unique characteristics of new computing systems will be achieved by the usage of a new, forth element of a fundamental sequence of electrical engineering elements: along with resistors, capacitors and inductors, invented in 1971 memristors will be used. A memristor can be defined as an electrical network passive element the resistance of which depends on the charge passed through it. After switching-off of voltage in the network a memristor does not change its state, i.e., it "memorizes" the last value of resistance. That is how the metastable On and Off states are realized. The further investigations demonstrated that the properties of a memristor enable to use it also as a switch, a ReRAM memory element. The effect of resistive switching (RS) or its variety, the bipolar effect of resistive switching (BERS), in the structures based on oxide compounds of a wide class including compounds of the strongly correlated electronic systems (SCES) class underlies these investigations. [1-8]. Academic and industrial directions of the investigations on ReRAM represent a wide range: the problem of materials, mechanism of RS, production, integration, etc. In this paper we present the review of the works in which memristor structures on the basis of high-temperature superconductors (HTSCs), critical analysis of these investigations conclusions and forecast of possible usage of their functional HTSC properties in structures formation and RS mechanism revealing are realized. Today no one theoretical model can fully explain the phenomenon of switching because of the lack of fundamental understanding of resistive switching process. In order to explain the acting principles exactly, one should have a comprehensive understanding of resistance switching mechanism on an atomic level, that is how conduction paths form and decay. Currently the study of oxide materials has two main directions: integration of new oxides with high permittivity to silicon technology and development of entirely new electronic devices. The occurring problem of reduction of power inputs in electronics and a number of other problems set the vector of the search for materials and devices with new functional properties. In this regard development of memory elements based on BERS effect permits to create new-generation elements. BERS reflects in the fact that in HTSC – normal metal heterocontacts the change in phase composition of SCES surface layer at nano-sized level occurs at certain polarity of the electric field. As a result, the metastable high-resistance (Off) and low-resistance (On) states of a heterocontact are realized, colossal electroresistance (CER) appears [3]. Colossal electroresistance (CER) is the ratio: CER = ($R_{off}$ -$R_{on}$)/ $R_{on}$, $R_{off}$ is resistance in the Off state, $R_{on}$ is resistance in the On state, it characterizes the memory effect

On the basis of physical processes that underlie devices with resistive switching they can be divided into three types [2]:

1. Structures which are underpinned by phase transitions from the amorphous to the crystalline state. (**Phase Change Memory technology**)

2. Structures based on redox processes in an electrolytic cell. As a result, clusterization of current paths, formation of thin conductive filaments between two electrodes occurs (**Ionic memory technology**).

3. Heterostructures based on strongly correlated compounds. (correlated electron random access memory (**CeRAM**).

*The first type* - phase memory is based on amorphous – crystalline state transition in a material (amorphous chalcogenide glasses, amorphous oxides of transition metals). It is related to energy release of electric field for heating of transition region. Current-voltage characteristics (CVC) of such transitions are unipolar. This problem is studied well enough and grounded theoretically. It is applied in practice.

*The second type* is RS effects related to electron transport in a solid electrolyte. At that several scenarios can develop. A new phase forms under the influence of the electric field in a percolation channel (electroforming process). A junction specific region, for example, Schottky barrier arises in a heterojunction near electrodes. The properties of the structure are determined by material properties, controlled by the electric field and modified by ion transport. Such a situation enables to methodically investigate the process of switching and making of a theoretical model for comparison with an experiment and to directionally design real memristors.

*The third type*. Today strongly correlated electronics is the most promising, but the least developed. Cuprates with high-temperature superconductivity and manganites with colossal magnetoresistance, transition-metal oxides, etc., are the typical representatives of strongly correlated materials. The electric and magnetic fields, pressure or light cause an electronic phase transition which includes big change in resistance (by several orders of magnitude) and is equivalent to the transition between metal and insulator. That is why it is expected that high-speed electronic devices with low energy consumption can be developed on the basis of such electronics.

Preceding the results of this review one can say that today active and numerous investigations, mainly with the second type of junctions, are carried out; and the effect of resistive switching is observed in a wide range of structures with a dielectric layer consisting of both simple oxides and complex compounds. In this review we will consider the studies of BERS in superconducting HTSC-based structures in which structures with BERS are obtained using functional properties of HTSCs. Several important factors should be emphasized.

*The first factor* is that parent HTSC compounds are antiferromagnetic Mott dielectrics, and **at *doping with oxygen metal-insulator (MI) transition occurs* [9] (Fig.1)** This circumstance is the *dominant functional* property of HTSCs for formation of heterostructures on their basis in order to realize BERS. As follows from, for example, a phase diagram at oxygen content between 6 and 6.5 YBCO is dielectric (let us denote this phase as $YBCO_6$), at oxygen content between 6.5 and 7 YBCO is superconductor (let us denote this phase as $YBCO_7$).

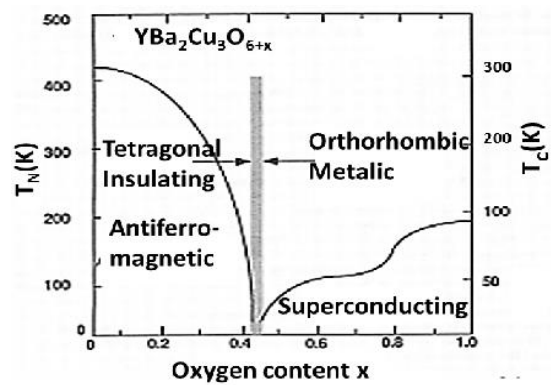

*Fig1. Phase diagram of YBaCuO [Source:9], $T_N$- Neel temperature, $T_c$- temperature of the superconductive transition. (Crosshatched area of two-phase state [9]).*

*The second factor* is that all the studied classes of superconductors are strongly anisotropic quasi-two-dimensional conductors with conductive copper (bismuth) – oxide layers (OL) divided by insulating charge reservoirs (CR) [9]. It is this circumstance that permits to regulate the type of conductivity by the electric field at a microlevel. And it should be noted that all the structures with BERS discussed in this review are performed for *c-oriented* single crystals and also for *c-oriented* epitaxial films.

*The third factor* is that all the junctions under study can be divided into the structures with vertical and planar geometry. The first type is symmetrical metal-insulator-metal (MIM) junctions with micro- or nano- dimensions. Planar structures are also MIM junctions with one electrode having the dimension significantly less than the general dimension of the structure. This can be both Sharvin-type microcontacts and sprayed electrodes. top electrode (TE). In [10, 11] it was demonstrated that in such planar structures the topology of electric field distribution E(x,y,z) considerably influences resistive switching. *Let us note that here everywhere we mean E current field $J=\sigma E$ which not always can be determined as E=V/d, as it is usually calculated, where V is voltage applied to the d-dimensional structure.* Let us emphasize the importance of this circumstance. In most works it is not analyzed how the electric field is distributed in a structure and how this distribution changes in the process of switching. In [10, 11] *the dominant role of inhomogeneous field distribution* in planar structures is underlined and it is shown that in this case c-regions with maximum electric field (critical regions) appear. As a result, *a percolation path in the form of a ring* is formed on the edge of the upper contact [10].

Finally, the essential condition for observation of the effect is the presence of a surface oxygen degraded layer for dimensions of the order of 10 nm with conductance different from the bulk one.

**$Bi_2Sr_2CaCu_2O_{8+y}$**

The first works [12-14] in which resistive switching in HTSCs was observed were made for the structures of a microcontact type (Sharvin point contacts) in BSCCO and YBCO. At that time a large experience in work with such contacts in the experiments on study of electron transport and electron-phonon interaction in metals and compounds was gained. Drawing on this experience in these works the results were obtained and the conclusions were drawn being important for understanding of BERS nature and HTSCs nature itself.

Today the analysis of all the investigations on HTSC memristor properties enables to conclude that the main functional property discussed above and oxygen loss on the surface of films and HTSC single crystals were exploited mainly for development of memristor structures. The first close structural investigations of BSCCO single crystals [14-16] determined that cleaving of the single crystal occurs along BiO planes, and in high vacuum it is stable during a month [17, 18]. In air the surface of cleavage degrades by oxygen for dimension of the order of 10 nm [15] (Fig.2).

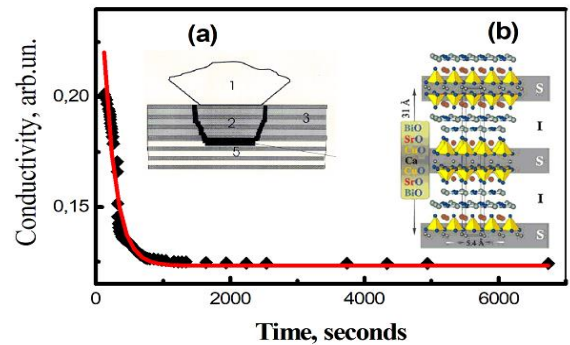

*Fig.2. Dependence conductivity of a point contact: single crystal cleavage – metal electrode on time. a) Contact scheme. 1-metal electrode,2-surface degraded region of a single crystal under the contact. 3- surface degraded region of a single crystal.4-boundary between a degraded region and a bulk part of a single crystal. 5-bulk non-degraded part of a crystal. b) The crystal structure of BSCCO.*

So, this surface layer determines the resistance of Ag/BSCCO microcontact: it is known [19, 20] that in point contact geometry of current spreading, heterocontact resistance is determined by volume of the order of d³ from the side of both contacting electrodes (d is contact diameter). The resistance of each electrode can be calculated from the ratio:

$R_i = \rho_i/2d$, (1)

where $\rho_i$ is the resistance of one of the electrodes. If on the surface there is one or several interlayers with different resistance $\rho_{sk}$, an additional contribution appears [20]:

$R_s = \Sigma^N_1 R_{sk}$, $R_{sk} \approx (\rho_i/2d)*\{(\rho_{sk}/\rho_i)-1\}*(z_k/d)$, (2)

From (2) colossal electroresistance will be determined by resistive properties and the dimension of a surface layer:

$CER = (\rho_i/2d)*\{(\rho_{sk}/\rho_i)-1\}*(z_k/d)/\rho_i/2d =$
$(\rho_{off}/\rho_{on})*(z_k/d)$ (3)

In [16] it was found out that normal metal – degraded $Bi_2SrCaCu_2O_{8+x}$ (BSCCO) surface interface has *diode properties*, and current transport through such an interface in BSCCO-based heterocontacts is carried out in a hysteresis way (Fig.3).

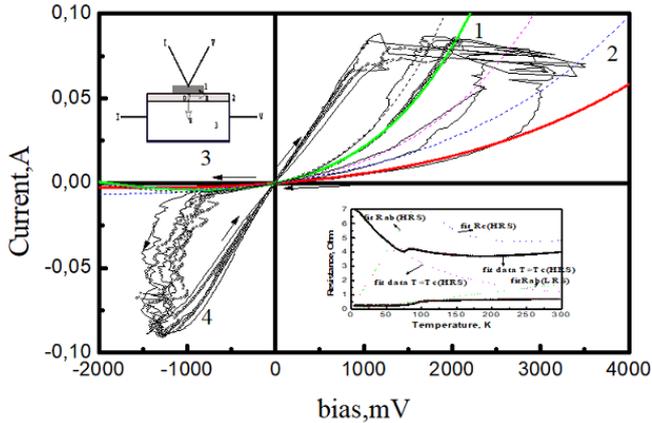

*Fig.3. Example of current-voltage characteristic of $Bi_2Sr_2CaCu_2O_{8+y}$ single crystal – Ag heterocontact with reproducible switching between the LRS and HRS states. Arrows show the directions of voltage sweeping. In this example the electric field - $\bar{E}(J)$ is directed from the surface deep into the crystal, when sweeping voltage is positive $\bar{E}(J)=J*\sigma$ (J is current density, $\sigma$ is conductivity. Contact scheme. 1-metal electrode, 2-surface degraded region of a single crystal under the contact. 3- single crystal. Inset: Temperature dependence of the HRS and LRS resistive states of Ag – $Bi_2Sr_2CaCu_2O_{8+d}$ single crystal heterocontact and approximation of the contributions of $R_{ab}$, $R_c$ to impedance [modified 16].*

A high-resistance branch of CVC with switching follows from the ratio (1):

$I(V) = I_o *(exp-(V/V_o)-1)$ (4)

Such dependence is typical for Schottky diodes [21].

As seen from the figure 3 switching in BSCCO is of multiplay nature. Today it is proposed to use this fact in the field of neurocomputing. Indeed, such memristor behavior offers a very simple solution to make artificial synapsis. A surface layer of HTSCs has carrier concentration less than volume. Temperature dependences of microcontacts in the metastable states were studied. As a result of significant anisotropy in c and ab directions, a point contact can be presented by an equivalent scheme of parallel connections $R_{ci}$ and $R_{abi}$. A surface layer under the contact is protected from degradation by the top electrode. That is why a surface layer in ab direction degrades more intensively, and the voltage applied to a contact will mainly drop on a surface degraded layer under the contact. It is significant electric field up to $10^5$ V/cm that stimulate carrier trapping to deep levels and dielectrization of the region under a contact. In the boundary of this region Schottky barrier appears. CVC of the appeared high-resistance phase becomes diode. At inverse voltage sweeping the region under a contact turns to a low-resistance metal state. As the analysis of the temperature dependence of contact resistance shows, in resistance there is a significant contribution of localization type proportional to $T^P$ where $p <-1$ [22]. Varying the external parameters—frequency and strength of the electric field applied to the heterocontact — different metastable states were actualized. These states may be use how electron memory.

Two important observations were made. The *first* one is that CVC of the structures demonstrating BERS can be described by a universal dependence of *Schottky diodes* (1). Such dependence was further discovered in other memristor structures made on film heterostructures of mesoscopic type. The *second* one is that by the study of an electronic structure of BSCCO surface layer by XPS methods it was shown that modification of the surface (degradation) leads to the change in O1s oxygen state. *This change can be related to vacancies in BiO layer, or, as they assume in calculations of electronic [18] BSCCO structure and in the study of X-ray photoelectron spectroscopy, oxygen is weakly bound in cuprate planes and at surface degradation oxygen vacancies appear there.* Such change in oxygen states was observed further both in structures with BERS based on HTSCs and other oxide compounds. This enabled to explain BERS by the effect of electric field on mobile vacancies, and formation of a percolation channel (filaments) – as the process of charge system redistribution in an active layer of structure.

In further works [23] on Ag/ BSCCO heterostructures the dynamics of Ag/ BSCCO heterostructures from one metastable state to the other one was studied. It was presented that rapid processes (of less than microseconds) and slow processes (of the order of ten seconds) are observed. They depend on voltage level and relate to oxygen electrodiffusion to vacancies (Fig.4).

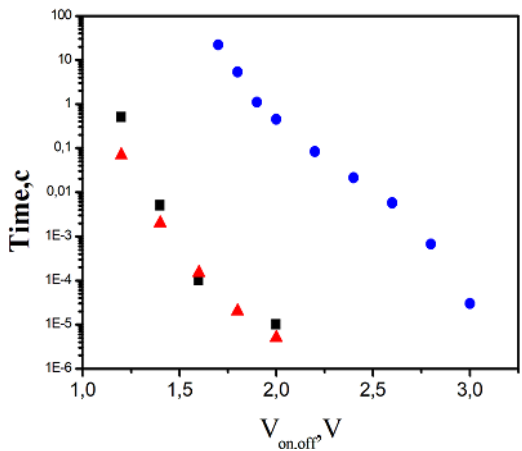

*Fig.4. Dependence of switching time on the pulse amplitude of voltage applied to $Bi_2Sr_2CaCu_2O_{8+y}$ single crystal – Ag heterostructure. Filled circles mean the process of junction from On to Off, filled squares and filled triangles mean the process of junction from Off to On.*

In [24] the effect of conductivity anisotropy and diffusion properties of a material on bipolar effect of resistive switching in BSCCO was analyzed by the method of numerical simulation. The results of numerical calculations showed that accounting of anisotropy of resistive and diffusion properties of heterostructure material leads to a considerable change both in electric field pattern and conducting channel form. The mathematical model permitted to calculate time dependences of heterostructure junction from a high-resistance (Off) to a low-resistance state (On) and inversely; it confirmed experimentally observed characteristics of resistive switching in heterojunctions based on $Bi_2Sr_2CaCu_2O_{8+x}$ high-temperature superconductor [23] having purely high conductivity anisotropy.

We should note the work [25] in which the role of electrodes in degradation of BSCCO single crystal surface layer and in effect of this process on RS is considered.

Thus, BERS in the structures based on c-oriented BSCCO single crystals are determined by a surface dielectric layer forming at MI transition with oxygen loss. This layer forms an *internal barrier* with a bulk superconducting metal and an *external barrier* with a metal electrode. As a result, *diode* properties of such a structure can be realized in a memristor device.

Works with mesas of BSCCO single crystals in the studies of internal Josephson tunneling deserve special discussion [26, 27]. In the works with mesas it was demonstrated that oxygen layers conducting copper can dope at current flowing through charge reservoirs in the process of carrier trapping. The authors named this phenomenon **floating gate concept** and assumed that it can be a general property of layered materials where the insulating charge reservoir layers are separated from the conducting planes. Furthermore, a new electrical doping method for Bi-2212 mesa structures was found out that at application of significant voltage to a mesoscopic structure it is possible to change (reversibly) normal electrical resistance, critical current, superconducting gap and critical temperature alloying has the same effect as doping by the change of oxygen content. Thus, it is assumed that charge transfer occurs inside crystal layers, and/or reordering of oxygen vacancies occurs. In continuation of these works it was discovered that there are two types of effects: even and odd in the *electric field [28,29]*. However, the question about the distinction between even and odd in the field effects remains unexplained.

**YBa$_2$Cu$_3$O$_{7-d}$**

YBCO system was also actively studied in memristor direction, first in microcontacts [13], and then in film heterostructures produced by a modern technique of lithography [30-31]. In investigations of BERS based on this HTSC system they made almost the same conclusions as for that based on BSSCO system. To be exact: a surface layer of the film is degraded by oxygen, and according to a phase diagram of YBCO system [9] YBCO$_{6.5}$ dielectric layer is formed on the surface, and it determines memristor properties of the structures formed of these compounds (Fig.5 a,).

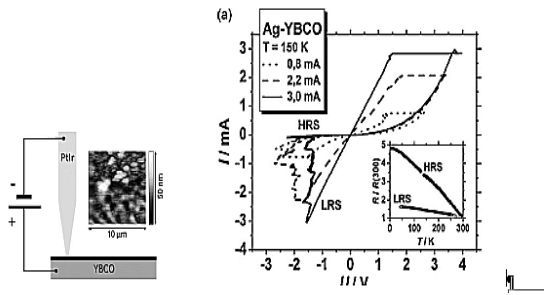

*Fig.5. a) Scheme of STM measurements. Inset is topology of YBCO film. b) CVC of Ag-YBCO contact in different current ranges and temperature dependence of the On and Off states (modified [30]).*

The authors of [32-33] report on resistive effect of switching in YBa2Cu3O6+x/ Nb-doped SrTiO3 heterojunctions. CVC of these heterojunctions show hysteresis which increases with decreasing temperature and oxygen content. CVC and temperature dependence of resistance [R (T)] of the interfaces of Au/YBa2Cu3O$_{7-d}$ were studied in different resistive states [34]. These states were obtained by resistive switching after N cyclic electrical pulses with different voltage amplitudes. CVC and R (T) dependences of different states are described within *Poole–Frenkel (P-F) mechanism* with layer trapping energy Et in a range of 0.06-0.11 eV. Et does not depend on the number of pulses and increases linearly with increasing amplitude of pulses voltage. The observation of P-F mechanism demonstrates the existence of depleted by oxygen YBCO layer near the interface. As a result, electric transport of such structures is determined by disorder degree, energy level of traps. In different temperature ranges nonlinear effects of electron transport are related to either P-F emission or hopping mechanism (VHR mode).

In [35] rectifying properties of YBCO/Nb-doped SrTiO3 structures were observed. A band diagram of the structure was determined from capacitive properties of these junctions (Fig.6).

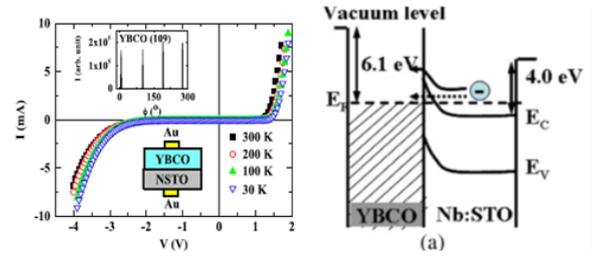

*Fig.6. Band diagram of YBCO/NSTO Schottky junctions (modified [35]).*

In [36] they carried out the investigation on electronic structure of the surface of c-oriented epitaxial YBCO films by the methods of X-ray photoelectron spectroscopy (XPS) and atomic-force microscopy (AFM). It was demonstrated that the surface layer (10-30 nm) is doped with oxygen from the metal (in film volume) to the dielectric state on the surface (Fig.7).

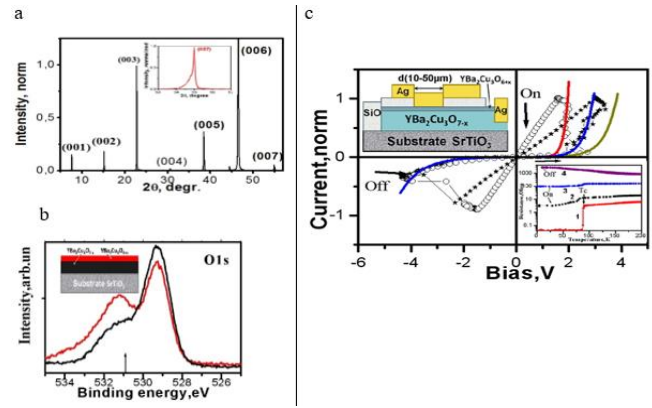

*Fig.7. a). X-ray spectrum of initial epitaxial YBCO films. "c"- parameter of the lattice "c" calculated by diffraction patterns that turned out to be c= 11.6935+-0.0003. On the basis on the data from [37] where a correlation between lattice parameter c and oxygen content in YBa$_2$Cu$_3$O$_{7-\delta}$ was found, a parameter $\delta = 0.14$ was calculated. A more detailed analysis showed that in addition to the main composition, the regions with smoothly decreasing*

*oxygen content to δ= 1.0 are present in the sample. In Fig.7a it is reflected in appearance of an extended "left arm" to the values of the lattice parameter up to 11.83 A. b) XPS spectrum of initial (red lines) and etched into a depth of the order of 30 nm (black line) YBCO films demonstrates oxygen spectrum in dielectric YBCO phase and in the phase of stochiometric composition after etching.c) Examples of resistive switchings in mesoscopic $YBa_2Cu_3O_{7-\delta}$/Ag heterocontacts obtained by lithography. Current is normalized to maximum value. In the upper left corner there is a scheme of the structure, in the lower right corner there are temperature dependences of the resistance of the initial film (1) along film plane direction "a" and the metastable states (2, 3, 4) along "c" direction. Depending on the value of flowing current different metastable states, from the dielectric (Off) to the metal (On) one, are realized. The solid lines indicate the CVC approximations by relationship (5). (modified [36]).*

Thus, YBCO-based structures are the unique objects to study BERS in HTSCs when an internal superconductor – doped dielectric barrier appears in heterostructures on their basis. Heterostructures based on epitaxial films of YBCO high-temperature superconductors, where BERS with realization of superconducting junction in the metastable On state (Fig.7c) are realized and studied, were developed by the method of photolithography. CVC with BERS is of diode nature. Current transport in metastable Off states of the studied geterostructures is approximated by the behavior of **two diodes switched on towards each** other. CVC of the structures evidences the presence of Rs – series spreading resistance and parallel jumpers $R_i$ that should be taken into account while modeling transport properties of the studied structures by the ratio:

$I_{off}$ = -$I_{01}$ *(exp((-eV)/$n_1$*kT)-1) +

$I_{02}$*(exp((eV-I*Rs)/$n_2$*kT)-1) +V/$R_i$         (5)

where $I_0$ = A *$T^2$ *S *exp (− **e*ϕ$_B$**)/ kT), A is Richardson constant, S is contact area, ϕ$_B$ is barrier height.

Approximation of CVC by the ratio of (5) is presented in Fig. 7 c. The transition from a high-resistance to a low-resistance state (and inversely) is determined by the formation (and decay) of a percolation channel on the branches "1-2", "3-4". As it was shown in the works, in planar junctions *a percolation channel is formed in the form of a ring*.

A common disadvantage of many studies on memristor structures is the absence of data on real distribution of the electric field in the structures containing defects of different types or several interfaces. It is distribution of the electric field which depends on the resistive properties of different junction regions that in its turn changes these resistive properties in the process of switching, determines the dynamics of resistive switching. In this direction one should mention the works [10, 12] where it was demonstrated by numerical methods that in planar structures the topology of electric field distribution significantly influences resistive switching. An essential condition for observation of the effect is the presence of a surface layer (depleted by carriers) for dimensions of the order of 10 nm with conductivity different from the bulk one. In this case inhomogeneous distribution of the electric field forms regions with heightened electric field (critical regions), a percolation path in the form of a *ring* is formed on the edge of the upper contact [12].

Thus, in the analysis of memristor properties of BSCCO- and YBCO-bases structures discussed above it is shown that BERS based on these compounds are realized as a result of HTSCs functional properties, i.e. transition to the dielectric state at oxygen loss. As a result, MIM structures, where one electrode is either a Sharvin microcontact or a film of silver of other metal evaporated through a stencil mask, are realized by different methods. The other electrode is HTSC single crystal or film. M (bottom) is the bottom electrode which is a material of stochiometric composition itself and also a metal. At that barriers modifying by the electric field in the process of switching appear in the top and bottom boundaries of the electrode.

**$Ba_{0.6}K_{0.4}BiO_{3-X}$**

As is known, the discussed above HTSC classes are hole-doped conductors. The role of carrier type turned out to be nontrivial in HTSC problem, and today it is actively studied both theoretically [38, 39] and experimentally. Mott antiferromagnetic dielectrics, the representatives of strongly correlated electronic systems (SCES), are parent compounds of HTSCs. Cation doping and stoichiometry change by oxygen *turns the system into a metal state with superconducting transition in T=Tc.*

Most HTSCs representatives: hole-doped (HD) compounds [3], $Nd_{2-x}Ce_xCuO_{4-y}$, $Pr_{2-x}Ce_xCuO_{4-}$, and also bismuthates: $Ba_{1-x}K_xBiO_{3-y}$ and $Ba_{1-x}Pb_xBiO_{3-x}$ – are a few representatives of electron-doped (ED) compounds in HTSCs [3]. Phase diagrams of HD and ED differ: the region of superconducting state existence is narrower in ED systems. It can relate to the fact that Fermi level in HD HTSCs gets to a broad band, and in ED HTSCs it gets to a narrow band [4] (Fig.8).

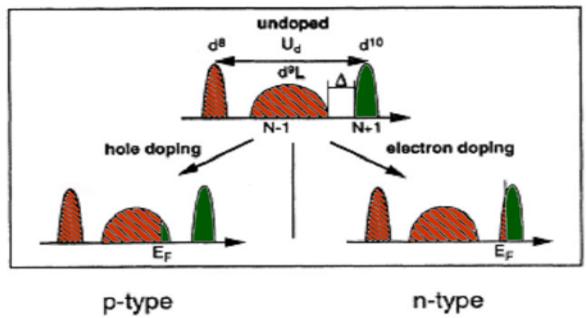

*Fig.8. Shematic representation of the influence of doping effect on Fermi level position in hole-doped and electron-doped SCES compounds (from D. Dagott // Rev. Mod. Phys. 66 (1994) 763).*

Physical origin of asymmetry between LSCO and NCCO is not only in the difference of oxygen – oxygen overlap value that controls derivative of Fermi surface, but also in difference of the value of charge transfer energy in these two structures. The last is of electrostatic nature: a material lacks for an electron (negatively charged apical oxygen that increases electrostatic potential on copper side).

The method of Hall effect measurement is the traditional method of carrier type study. The BERS in structures based on strongly correlated electronic systems turned out to be sensitive to carrier type; they are the original method of control or determination of carrier types in such systems as HTSCs and doped manganites.

BKBO system was investigated only in one work [40], where BERS in heterojunctions based on Ba0.6K0.4BiO3-y single crystals was observed, and it was shown that the effect has an opposite sign in voltage as compared to an analogous effect observed in the structures based on hole-doped systems: switching to the metastable high-resistance phase occurs when HTSC single crystal is in negative potential in respect to a normal electrode. In this case current field E (J=σ*E) is directed to the surface in contrast to the switching observed in hole-doped compounds (Fig.9).

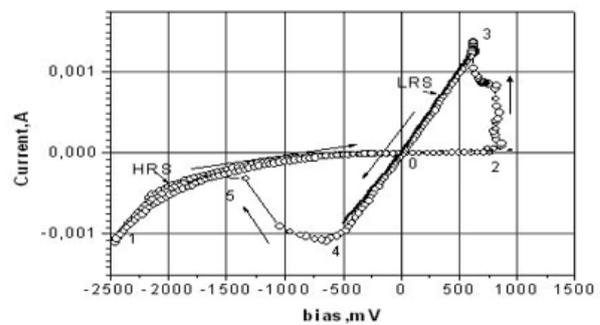

a

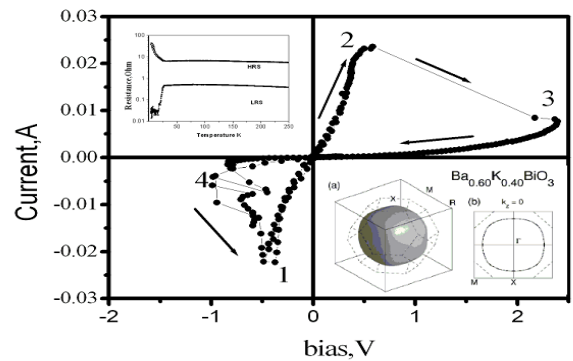

b

*Fig.9 a) Example of CVC of Ag film/BSCCO junction with BERS. b) ) Example of CVC of Ag film/BKBO junction with BERS. In left inset there is the example of temperature dependence of resistance of two metastable contact states: the low-resistance "LRS" ("On") state with junction to a superconducting state and the high-resistance "HRS" ("Off") state. In right lower coner- Fermi surface of cubic Ba1-xKxBiO3 for x=0.67.*

It should be noted that today in this class there are no investigations on the effect of oxygen nonstoichiometry on BKBO properties like those carried out in BSCCO and YBCO. However, from the work on tunneling in Ag/BKBO junction it is possible to judge by implication that the surface layer of a single crystal is dielectric served as a tunneling interlayer [42]. Tunnel studies in this work demonstrated that in this class of compounds superconductivity is well described by electron-phonon interaction. Such junction is classic M(top)-I-M(bottom) structure. As it can be seen in Fig.9 from CVC with switching, such structures have diode properties with inversion of switching polarity as compared to hole-doped structures.

### $Nd_{1.86}Ce_{0.14}CuO_4$

NCCO is also the representative of electron-doped high-temperature superconductors. In [43, 44, 45] detailed investigations of BERS in various epitaxial NCCO-based structures were carried out. They found that c-oriented epitaxial NCCO films represent a structure with few tens nm thick phase of oxygen-deficient $Nd_{0.5}Ce_{0.5}O_{1.75}$ (NCO) oxide intergrown on the surface. It permitted to produce two types of heterojunctions: NCCO/ NCO/ Ag and NCCO/ Ag on the surface of NCCO films before and after ion etching. Fig.10, 11 show the results of XSA and XPS studies of initial and etched into 30 nm NCCO films. Thus, we obtained two types of heterojunctions for NCCO films: a) with surface buffer layer of $Nd_{0.5}Ce_{0.5}O_{1.75}$ oxide (Fig10.a), see below the data of X-ray investigations, b) without $Nd_{0.5}Ce_{0.5}O_{1.75}$ oxide layer which was removed by ion etching up to a main NCCO phase under photoemission control (Fig.10b).

Figure 10 d demonstrates the examples of CVC of $Nd_{1.86}Ce_{0.14}CuO_{4-y}$/Ag heterojunction with resistive switching at quasi-stationary voltage sweeping ($f=10^{-3}$Hz). The positive direction of electric current field at voltage sweeping from the bottom to the top electrode was chosen. In the heterocontacts formed directly on the etched surface of NCCO films the effect was insignificant in value, at that switching pattern corresponded to electron-doped systems. One can understand this fact considering weak effect of oxygen as a doping element on the properties of $Nd_{2-x}Ce_xCuO_{4-y}$. Today this circumstance is proved in a numerous of studies.

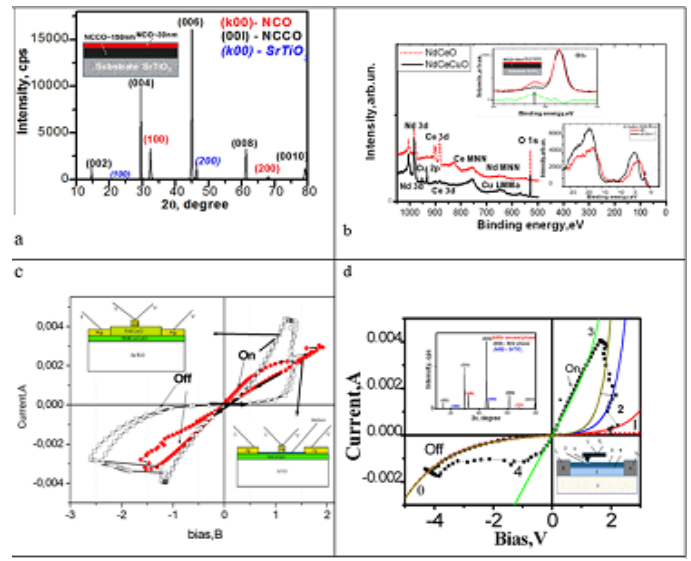

*Fig.10. X-ray diffraction analysis (a) and X-ray photoemission spectra (b) of initial (red lines) and after ion etching NCO/NCCO films (etching depth is 25-30 nm, black lines) (c,d). Examples of CVC of NCCO/interface/Ag heterojunction with resistive switching at quasi-stationary voltage sweeping ($f=10^{-3}$Hz).*

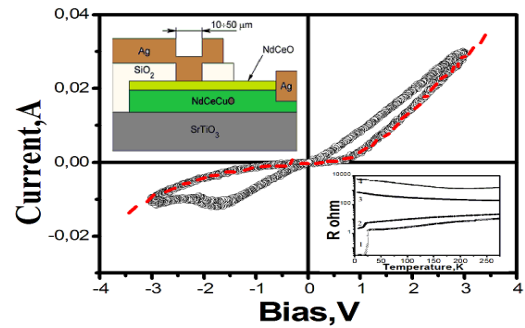

*Fig.11. Example of CVC of lithography-formed structure metastable states-exhibiting NCCO/NCO-based heterostructure. The solid lines indicate the CVC approximations by ratios 5 (see the comments in the text). Lower right corner - temperature dependence of resistance of initial superconducting NCCO film (1) and metastable ON (2) and OFF (3,4) states. In left corner: schematic of lithography-formed structure representation.*

Basing on the experimental results, the model of resistive switching mechanism in the studied heterostructures is proposed in the work [45].

Figure 12 demonstrates the equivalent schemes of the main CVC branches of heterocontacts with switching and model calculations of formation and decay of a percolation channel in the model "*critical region of the electric field*" considered in work [12 and references]. *According to this model, the change in resistive properties of the heterostructure described above at application of voltage of different polarity to its contacts is associated with formation (or decay) of a "conducting channel" through a dielectric layer. At that it is assumed that the change in specific conductivity occurs in those parts of the specified layer where electric field exceeding some critical value is achieved. Mathematical formulation of the model is based on equation of current spreading: $\nabla(\sigma\nabla\varphi) = 0$ which enables to determine distribution of electric field potential $\varphi$ by numerical methods, and then also to determine distribution of electric field by distribution of specific conductivity $\sigma$ in a heterostructure.*

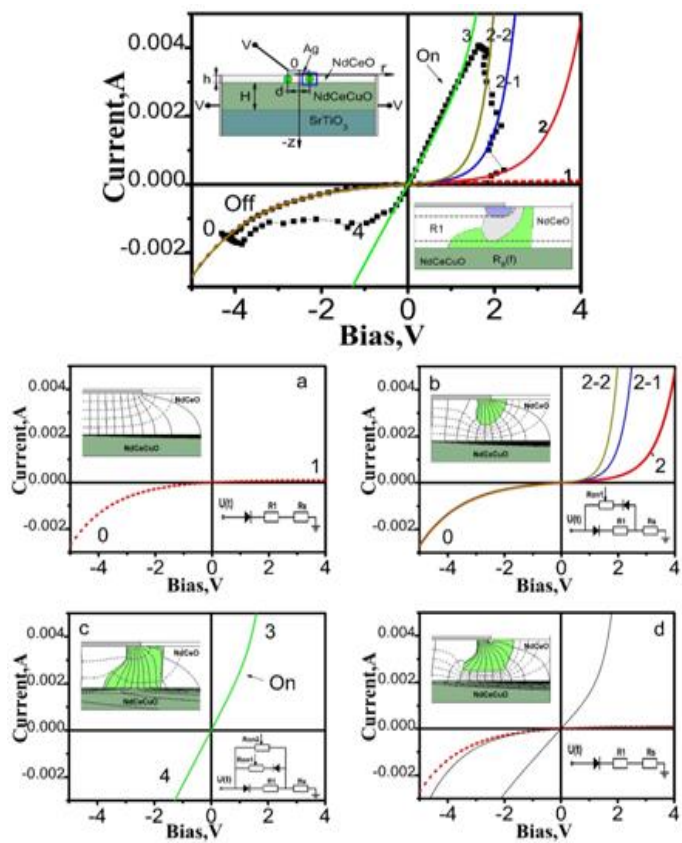

*Fig.12. Equivalent schemes of the main CVC branches of NCCO/NCO/Ag heterocontacts with switchings and model calculations of formation and decay of a percolation channel in the model "critical region of the electric field" considered in work [45]*

The initial state of heterojunction with BERS – diode with Schottky barrier (CVC branch 01). At achieving certain value of voltage the first region, where critical electric field is achieved, appears on contact edge. As a result of Poole–Frenkel effect, in this region the number of carriers increases, carrier tunneling begins, heterostructure properties change, and resistance decreases (branch 2-3). As voltage increases, a channel with increased conductivity is forming. When the channel of increased conductivity reaches an inner NCO/NCCO boundary, a diode directed oppositely to the first one is formed. At that electric field distribution changes. Simultaneously with this rapid process oxygen ions diffuse to vacancies. The number of carriers increases, diodity disappears, the electrode transits to the ohmic state. Thereby, the formation of *a channel in the metastable On state* in the form of *a ring* is completed. Then it is the low-resistance state (branch 3-4). In branch 4-0 a junction to the high-resistance state occurs. The start of the process of conducting contact decay is reverse electrodiffusion of oxygen and vacancies, carrier trapping, redistribution of vacancies and recovery of the Off state. Thus, resistive switching in the studied structures is the result of three processes having different characteristic times: 1) change in carriers number, apparently due to accumulation of minority carriers in the depletion region of Schottky barrier and in film volume; 2) field effect, formation of domains with increased carrier density and change in electric field pattern related to the first and the second processes; 3) oxygen electrodiffusion that forms resistive properties of an interface and forms a parallel resistive channel. The effect of frequency within the framework of the *parallel diodes model* comes down to the conclusion that with frequency the number of carriers changes and a parallel channel fails to form.

Thus, resistive switching in the investigates structures is controlled by several factors: change in resistive state of normal metal – oxide interface influenced by AC voltage, formation of parallel channels of carrier passing, accumulation and relaxation of minority carriers in the depletion region close to electrodes and in NCO film volume, change in electric field pattern as a result of change in resistive properties of various regions in an active heterostructure part, oxygen electrodiffusion to vacancies that fixes the final On and Off states in different CVC branches of a heterocontact. Apparently, electrodiffusion is the slowest considered process; it limits the frequencies at which BERS is observed. The dominant factor for switching is the presence of oxygen-deficient NCO phase epitaxially intergrown on NCCO surface, geometric dimensions of a structure and the existence of an internal barrier on NCCO/NCO boundary.

**Conclusions**

Thus, from the investigations of BERS in HTSCs it follows that:

1. On the surface of HTCS single crystals and epitaxial films oriented (001), as a result of degradation by oxygen in charge reservoirs a dielectric layer forms at a depth of the order of 10 nm. This fundamental property permits to use MIM heterojunctions based on perovskite compounds of HTSC type in development of memristors based on resistive switching.

2. Switching polarity and effect size in electron-doped and hole-doped reflect the nature and degree of oxygen doping of these compounds.

3. Current transport in metastable high-resistance states of the studied hetrostructures is approximated by the behavior of *two diodes switched on towards each other*.

4. Switching in HTSC-based structures is of multilayer nature. Today it is proposed to use this circumstance in the field of neurocomputing.

**Prospects of usage of HTSC-based memristors in applications and search for new mechanisms of strongly correlated nature to realize new-generation memristors**

Today strongly correlated electronics is the most promising, but the least developed. Cuprates with high-temperature superconductivity and manganites with colossal magnetoresistance, transition-metal oxides $VO_2$, NiO and others are the typical representatives of strongly correlated materials. The electric and magnetic fields, pressure or light cause an electronic phase transition which includes big change in resistance (by several orders of magnitude) and is equivalent to the junction between metal and insulator. That is why it is expected that high-speed electronic devices with low energy consumption can be developed on the basis of such electronics (ITRS) [46]. In the latest editions of the International Technology Roadmap for Semiconductors (ITRS) a new memory class "Mott memories" is discussed; it is a new mechanism of resistive switching based on Mott junctions (MIT). They can indeed undergo various kinds of insulator-to-metal transitions (MIT) in response to different external perturbations like pressure, temperature, and electronic filling [47].

1) Bandwidth controlled MIT. This MIT corresponds to the crossing of the Mott transition line induced by tuning the correlation strength U/W, where U is Coulomb repulsion, W is band width. This can be achieved by applying an external pressure which enhances the orbitals overlaps and increases thus the bandwidth W.

2) Temperature controlled MIT. This MIT, is driven by of U/W around $\approx 1.15$. This MIT occurs between a low temperature metal and a high temperature insulator which strongly contrasts with the more usual transitions from a low- T insulator to a high- T metal,

3) Filling controlled MIT. This MIT, occurs when the band filling deviates from half filling.

However, the role of strongly correlated effects in resistive switching of the structures on based-HTSC is still unobserved. The problem is that these compounds also contain oxygen ions. Oxygen vacancies play a impotant role in formation of the ground state of SCES. That is why it is difficult to distinguish if the effects of resistive switching in these structures are related to ion migration and modification of barrier properties in the electric field or the effect of the electric field on oxygen vacancies stimulates metal – insulator electronic transition. The observed in the works under discussion effect of resistive properties of HTSC – normal metal heterojunctions influenced by the electric field is the evidence in favor of a so-called filamentary model with electroforming. Recently several investigations in the field of technologies of non-filamentary non-volatile memory based on metal – insulator Mott transition in nickel oxide and other transition-metal compounds have appeared [48, 49]. It is assumed that electroforming process proves the filamentary mechanism of RS. However, in works [48] it is shown that in primary state NiO can be produced as metal that is achieved by doping of nickel carbonyl with ligands. At that electroforming is not required for conductivity to achieve initial metal state. Metal – insulator transition is carried out in a central region, at that any state can last for as long as necessary. This peculiarity is basic at considering CeRAM as a candidate for the role of volatile memory. Special requirements are imposed to barrier layers in structures: they provide ohmic contact with an active material, and remove any barriers related to Schottky effects caused by outer electrodes and any undesirable surface states from active material – electrodes boundary.

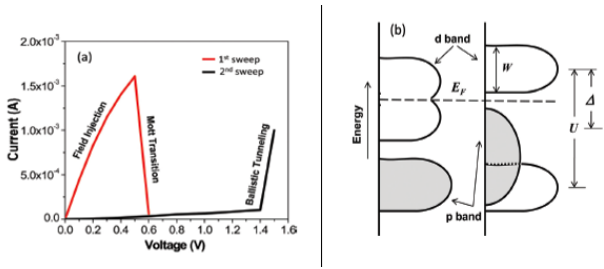

*Fig.13. a) CVC of CeRAM device based on the principles of electron correlations. b) Band diagram of NiO in the On and Off states where U is Coulomb repulsion, W is band width, $\Delta<U$.*

From these points of view all heterojunctions based on HTSCs do not exhibit such effects. On the other hand, layered HTSC-materials demonstrate well pronounced 2D-properties of current carriers in macroscopic 3D-crystals. Standard HTSC-materials contain conducting $CuO_2$ layers with oxygen pyramids (YBCO, BSCCO) or octahedrons (LSCO), while optimally annealed NCCO crystals contain $CuO_2$ layers without apical oxygen atoms, i.e. $CuO_2$ layers form quasi-two-dimensional (2D) planes. That is why HTSC single crystals can be considered as selectively doped system of quantum wells ($CuO_2$ layers) divided by the barriers doped by buffer layers. Relying on these quantum fundamental properties of HTSCs it is possible to say that they are excellent objects for construction from them memristor structures basing on new principles of strongly correlated electronics (using, for example, floating gate concept) or using the model of the work [50] where it is assumed that the reason of BERS is in the nature of Mott junction in normal metal – SCES interface where a two-dimensional structure metal/ band insulator/ metal/ Mott insulator / SCES is formed (Fig.14).

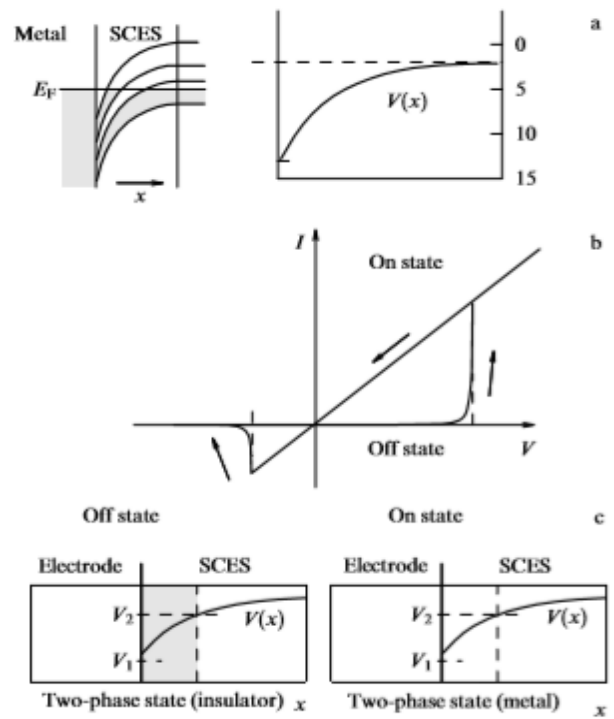

*Fig.14. a) Schematic diagram of the potential distribution over the metal/SCES interface, b) CVC of a metal SCES structure in the model .c) The off and on states of a metal –SCES structure [50].*

The mechanism of CER has been proposed here does not assume any interface states but attributes the large nonlinearity of the I-V characteristics to interface Mott transition, where a layer of Mott insulator blocks the current. In dimensions higher than one, the transition is first order, and the width of the Mott insulator layer depends on how the voltage is changed. This explains the hysteretic behavior of the I-V characteristics.